\documentclass[aps,prb,showpacs]{revtex4}

\usepackage{amsmath}
\usepackage{amsfonts}
\usepackage{graphicx}
\usepackage{color}

\newcommand{\be}{\begin{equation}}
\newcommand{\ee}{\end{equation}}
\newcommand{\bea}{\begin{eqnarray}}
\newcommand{\eea}{\end{eqnarray}}
\newcommand{\bc}{\begin{center}}
\newcommand{\ec}{\end{center}}

\newcommand{\fues}[1]{\left(#1\right)}
\newcommand{\Sh}{Schr\"odinger{ }}

\begin{document}

\title[]{Magnetic field barriers in graphene: an analytically solvable model}

\author{Enrique Milpas }
\email[ Email:]{saplim@fisica.unam.mx} 
\author{Manuel Torres}
\email[Email:]{torres@fisica.unam.mx} 
\affiliation{Instituto de F\'{\i}sica,
Universidad Nacional Aut\'onoma de M\'exico,
Apartado Postal 20-364,  M\'exico Distrito Federal 01000, M\'exico}
\author{Gabriela Murgu\'ia}
\email[Email:]{murguia@ciencias.unam.mx} 
\affiliation{Facultad de Ciencias,
Universidad Nacional Aut\'onoma de M\'exico,
Apartado Postal 21-092,  M\'exico Distrito Federal 04021, M\'exico}

\date{\today}
\pacs{72.80.Vp,73.21.-b,71.10.Pm,  03.65Pm}

\begin{abstract}

We study the dynamics  of carriers in  graphene  subjected to an inhomogeneous magnetic field. 
For a  magnetic field with  an hyperbolic profile  the corresponding Dirac equation can be analyzed 
 within the formalism of supersymmetric quantum mechanics, and leads to   an exactly solvable model. 
We study in detail the bound spectra. For a narrow barrier the spectra  is characterized  by a few bands, except for the zero energy level that remains degenerated. As, the  width of the barrier increases  we can track  the bands  evolution  into the degenerated Landau levels. In the scattering  regime  a simple analytical formula is obtained for the  transmission coefficient,  this result  allow us to identify the resonant conditions at which  the barrier becomes transparent.  

\end{abstract}

\maketitle

\section{Introduction}
The discovery of graphene \cite{novo:666,novo:197,zhan:201}, a single layer of carbon in a honeycomb lattice has generated a lot of 
excitement, due to its unique electronic properties \cite{kats:20,novo:1379,novo:177} and its potential application in electronic devices. 
Electrons in graphene are described by a   massless two  dimensional relativistic Dirac equation  \cite{wall:622,seme:2449,castro:109}, that yields a  gapless linear spectrum  near  to the $K$ and $K^\prime$ points of the first Brillouin zone.  Graphene exhibits a variety of pseudo relativistic phenomena, 
providing an unexpected  connection between condensed matter physics and  quantum-relativistic phenomena. Among others we can cite:
 the   Zitterbewegung and its relation with  the minimal electrical conductivity at vanishing carrier concentration \cite{novo:197,zieg:233407,castro:109}, the unconventional quantum hall effect \cite{novo:197,zhan:201,gusy:146801},    and the Klein tunneling \cite{kats:620,chei:041403,kats:157,youn:222}. The Klein tunneling effect has important implications  for the future design of graphene based electronic devices, because  massless Dirac fermions cannot  be effectively  confined  by electrostatic barriers, in particular for normal incidence the barrier becomes complete transparent \cite{chei:041403}.

 Some schemes have been proposed in order  to avoid the obstacle  that represents the Klein tunneling, in order  to   confine electrons  in graphene based structures. An interest proposal refers to the use of inhomogeneous magnetic fields
 that produce magnetic barriers \cite{mart:066802}. Previous studies have consider the cases of single  \cite{mart:066802}, double\cite{oros:081403} or multiple magnetic barriers  \cite{masi:235443, anna:1918}. All of these refer to square magnetic barriers with sharp edges. 
 In this paper  we consider a magnetic  barrier in which  the edges are smoothed out.   We select a
magnetic field with  an hyperbolic profile.   We show that the corresponding Dirac equations can be analyzed 
 within the formalism of supersymmetric quantum mechanics, and leads to an exactly solvable model. 
We study in detail the bound spectra. For a narrow barrier the spectra displays a series of bands separated by gaps,  as the 
width of the barrier increases the bands evolve into the degenerated Landau levels. In the scattering  regime  a simple analytical formula is obtained for the 
transmission coefficient,  this result  allow us to identify the resonant conditions at which  the barrier becomes transparent.

The use of inhomogeneous magnetic fields has received considerable attention both in the experimental \cite{lee:1,vancu:5074,carm:3009,ye:3013,novo:233312} and theoretical  \cite{lee:1,peet:15166,matu:1518,hand:161308}
study of two-dimensional electron gases (2DEG) in semiconductor heterostructures.
Various configurations of local inhomogeneous magnetic fields   have been created and studied, using microfabricated ferromagnetic and superconducting structures deposited on top of a  2DEG. Interesting transports phenomena have observed, among others: magnetoresistance and commensurable oscillations,   anomalous transport along special e.g., snakelike  trajectories,  etc.
 Although there exist  to date no experimental realization  of similar configurations in graphene, they should be produced in the   near future. The configuration for the hyperbolic magnetic field, see below  Eq. (\ref{magB}), provides a good approximation to the shape of the magnetic barrier produced by a ferromagnetic film deposited in a 2DEG  \cite{vancu:5074}. Hence we expect, that apart of its  intrinsic theoretical interest,  the  results obtained in this work will  be useful in order to analyze the confinement by  magnetic barriers in graphene samples.

The paper is organized as follows. In  section II we study  the model for Dirac fermion  dynamics in graphene when the system is subjected to an inhomogeneous magnetic field. We show that the system can be analyzed within the formalism of supersymmetric  quantum mechanics. The  effective potentials and the explicit analytical solution for the wave function are discussed. In section III we study in detail the bound spectra and analyze it corresponding  degeneracy, both for a narrow barrier, and also in the limit in which the barrier width becomes comparable to the size of the system. In section IV the dispersion regime is analyzed in detail.  In section V the conclusions are presented.

\section{Graphene in an inhomogeneous magnetic field}

\begin{figure}
\includegraphics[width=8cm]{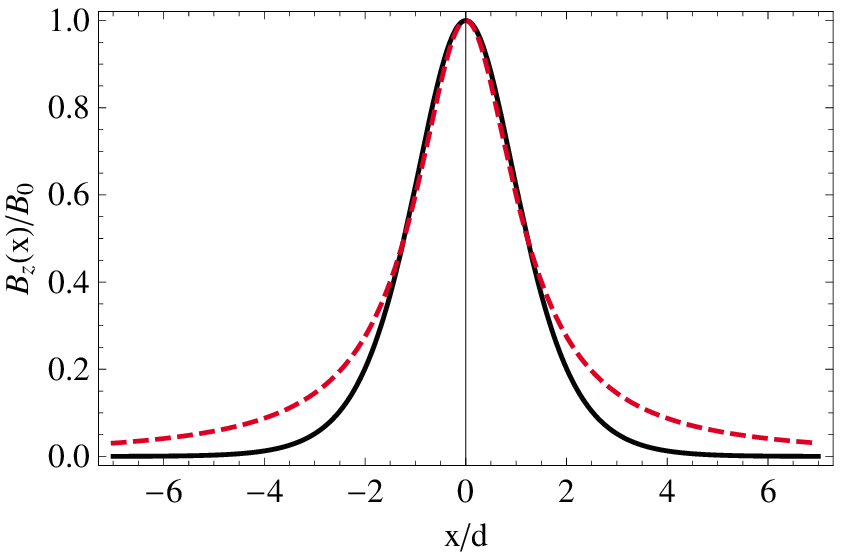}
\caption{(color online) Magnetic field profile for the hyperbolic expression in Eq. (\ref{magB}) (solid line) as compared to the 
profile  $B_z(x) = -( \mu_0M/4\pi ) \ln [(x^2 + k^2)/(x^2 +(l+h)^2)]$, \cite{vancu:5074} produced  by a ferromagnetic strip with  a magnetization  $\mu_0 M$,  thickness $l$ and  separated a distance  $h$ from the two dimensional system (dashed line). The parameters are selected in such a way that the peak and  half width of both profiles have the same value.}
\label{figureB(x)}
\end{figure}

We focus on  an  electron in a single  graphene layer subject to an inhomogeneous perpendicular  magnetic field, which varies along the $x$ direction. In order to  study an smooth magnetic barrier  we select a profile given by 
  \be \label{magB}
    \mathbf{B}  = B_0 Sech^2 \left( \frac{x }{2 \, d}\right) \,  \boldsymbol{e}_z  \, , 
    \ee
here  $ \boldsymbol{e}_z  $ is the unit vector normal to the  graphene plane. 
  This  expression for  the  magnetic field presents several advantages: 
$(i)$ We obtain an analytically  solvable model, that allows us to analyze in detail the bound spectra, as well as the  transmission  through the magnetic barrier. $(ii)$ In order to have conditions that are physically  relevant to the study of  graphene, we need a  
magnetic field $B(x)$  that varies slowly  on the scale of the graphene lattice spacing, $a = 0.246 \, nm$.  Selecting 
$ a \ll d$,   we observe  that both the half-width ($\Lambda \approx   3.25 \, d)$ and the edge smearing length  $ \vert (1 /B) (d B / dx) \vert \sim d$ of the magnetic barrier    satisfy the required conditions. $(iii)$ As shown in  Fig.(\ref{figureB(x)}) the magnetic field  in Eq. (\ref{magB})  provides a good approximation to the shape of the magnetic field barrier  that is produced  by a ferromagnetic film  deposited on the top of a two dimensional system \cite{vancu:5074}. Thus the present formalism could be useful to analyze a  similar arrange of  inhomogeneous  magnetic barriers  in graphene samples.

 The gauge can be  selected in such a way that the vector potential 
is written as $\mathbf{A } = \hat{e}_y \, A(x)$ with 
 
\be \label{magA} 
   A(x) = 2 \, B_0 \,  d \,  tanh \left( \frac{x }{2 \, d}\right)  \, .  
\ee
Notice that if we consider  that the system is confined within a square box of area $ L \times L$,  then in the  
 $d >> L$  limit,  the magnetic field can be consider homogeneous, and the vector potential reduces to the  
 Landau gauge expression  $\mathbf{A } = B_0\, (0,x,0)$. 
 
 On low energy scales the dynamics of quasiparticles in graphene is described by two independent 2+1 dimensional Dirac equations, the equations remain decoupled in the presence of smoothly varying magnetic field.  The resulting 
 time-independent  Dirac equation  describing low energy excitations around the 
 $K$  point in the Brillouin zone is written as

 \be \label{hamilt}
H  \,   \psi (x,y)\,= \,  v_F  \vec{\sigma} \cdot \left[ \mathbf{p} + e \mathbf{A}(x)   \right] \,   \psi (x,y)\,= \,  E  \psi (x,y)    \,, 
 \ee
 here the Fermi velocity is $v_F  \approx  c/300$, $ \mathbf{p} =  - i \hbar \nabla  $ is  the momentum operator and the isospin Pauli matrices $\sigma_i$ operate in the spinor  $  \psi(x,y) = \left(\psi_A \, ,\, \psi_B \right)^T$,   that represent the  electron amplitude on two sites ($A$ and $B$) in the unit cell of the graphene lattice.     
 Taking into account the translational invariance along the $y$ direction we seek solution of the 
 form  $\psi_A  = exp(i{ k_y \,  y}) \psi_+$ and $\psi_B  = exp(i{  k _y \,  y}) \psi_-$.  
  The Dirac equation  yields the coupled equations

   \bea \label{eqaco1}
   \Delta   \,  \psi_+ (x)   &=&     \left( - i \, \frac{\partial}{\partial  {\tilde x} } -  i W(x)  \right) \psi_- (x)   \,  ,\nonumber \\
   \Delta   \,  \psi_- (x)   &=&     \left( - i   \, \frac{\partial}{\partial  {\tilde x}} +  i W(x)  \right) \psi_+ (x)   \, ,
 \eea
with the magnetic length  defined as $l_B =\sqrt{\hbar /e B_0}$,  ${\tilde x} = x/l_B$, and  $\Delta = E \, l_B  /\hbar v_F$.   The function $W(x) $ is given by
  
  \be \label{superpot}
 W(x)    \, = \,  l_B  k_y + \frac{e l_B A(x)}{\hbar}    \, = \,   l_B  k_y  +  \frac{2   d}{l_B}  \tanh \left( \frac{  x }{2 \, d}\right)
   \, . 
 \ee

   Combining the two equations in  (\ref{eqaco1}) we obtain the decoupled equations
  \be \label{eqSheff}
 H_{\pm} \, \psi_{\pm} (x)  = \left(  -    \frac{d^2}{d   {\tilde x}^2 } + V_{\pm}  \right) \,  \psi_{\pm} (x) =   \Delta^2   \,  \psi_{\pm} (x)   \, , 
 \ee
 where the effective potentials $V_{\pm}$ are given by 
 
  \be \label{effV}
 V_{\pm}  (x) =     W^2 \pm \frac{d W }{d {\tilde x}} =  \left[  \left( l_B k _y + \frac{2 d}{l_B}   \, \tanh \left( \frac{ x }{2 \,  d}\right)   \right)^2   \,  \pm \, 
  \,  sech ^2 \left( \frac{ x }{2 \,  d}\right) \right] \, . 
 \ee

Both the shape and the depth of the effective potentials depend on  the  transverse momentum  $k_y$. 
As seen  in  Fig.(\ref{figureV(x)}), depending on the values of $k_y$ and $d$,  the effective potentials can assume the form of   potential wells or  steps. In  next sections the bound  state  spectrum and 
scattering properties   will be analyzed in detail.

 It is interesting to point out that the Dirac equation in the presence of an external magnetic fields posses a formal structure that  can be analyzed within the formalism of  supersymmetric quantum mechanics (SUSY-QM) \cite{coop:267,coop:0,kuru:455305}. The potentials $V_+$ and $V_-$ in (\ref{effV})  are  known as  the super-partner  potentials, they  are obtained from  the  superpotential function $W(x)$ by the  relation
 in (\ref{effV}). The explicit expressions for  $V_{\pm}$  for  the gauge potential in (\ref{magA}) 
 are identified as the Rosen-Morse II  potentials \cite{coop:267,coop:0} and  the  corresponding \Sh equations   are exactly solvable.  There are important  property of SUSY-QM  that  relate the spectrum and  eigenfunctions of the effective hamiltonians of $H_+$ and $H_-$ in Eq. (\ref{eqSheff}). In particular, except from the ground state,  $H_+$ and $H_-$ have the same spectrum for $\Delta^2$.
 
 It is convenient to define the operators 
 
 \be \label{opeL}
 L^{\pm} \, = \,  -  i \frac{d }{d {\tilde x}}  \pm i \, W(x) \, .
  \ee
 In terms of these operators the relation between  the upper and lower spinor components in  (\ref{eqaco1}) simply read 
 
   \be \label{eqaco2}
    \,  \psi_+ (x)   =    \frac{1}{  \Delta }  L^-  \psi_- (x)  \, , \hskip1.5cm \,
      \psi_- (x)   =    \frac{1}{  \Delta }  L^+ \psi_+ (x)  \, . 
 \ee
  Let us   introduce  the dimensionless variable   
  \be \label{varxi}
\xi = \frac{1}{1 + exp(x/ d) }\, , 
  \ee
 that varies from $0$ to $1$,   as $x$  goes from  $\infty$ to  $-\infty$.   
  In the new variable equations (\ref{eqSheff}) become
 
  \be \label{eqSh2}
\left[\frac{d^2}{d\xi^2} + \frac{1 - 2 \xi }{\xi (1 - \xi)} \frac{d}{d\xi}   +  \left(  \frac{d}{l_B} \right)^2 \frac{\Delta - \left[ k_y l_B + 2 d/l_B (1 - 2\xi) \right]^2  \pm  4 \xi(1 - \xi) }{\xi^2 (1 - \xi)^2} 
\right]   \psi_{\pm} = 0 \, .
  \ee
   These equations have the asymptotic solutions:  $  \psi_{\pm} \sim \xi^\rho$ for $\xi \to 0$ ($x \to \infty$) and    $\psi_{\pm} \sim (1 - \xi)^\sigma $ for $\xi \to 1$ ($x \to -\infty$), where the asymptotic behavior is determined by 

\be \label{coefas}
    \rho  \,  =     \,  \frac{d}{l_B}  \sqrt{\left( l_B \, k _y   + \frac{2  d}{l_B}    \right)^2  -  \Delta^2} \, ,  \hskip1.5cm 
       \sigma  \,  =     \,   \frac{d}{l_B}   \sqrt{\left( l_B  \,  k _y  - \frac{2  d}{l_B}    \right)^2  - \Delta^2}  \, . 
 \ee
In order to obtain consistent solutions, we recall that besides solving the effective \Sh  equation in  
(\ref{eqSh2}),  the wave function components are interrelated by the Dirac equation via (\ref{eqaco2}). Then,  one  can  consider the following options:
(a)  Equation (\ref{eqSh2}) is solved for  the lower component $ \psi_{-} $,  the corresponding upper component $ \psi_{+} $ is obtained from the first relation in  (\ref{eqaco2}).
(b) Equation (\ref{eqSh2}) is solved for  the upper component $ \psi_{+} $,  and the lower component 
is obtained from the second  relation in  (\ref{eqaco2}).
 We consider the first option, the second option gives equivalent solutions, except for the $n=0$ state. 
Taking into account the asymptotic behavior, we propose an  ansatz  of the form 
  $ \psi_{-} (\xi) =   \xi^\rho  (1 - \xi)^\sigma f(\xi)$, substituting in Eq. (\ref{eqSh2}) we find that $f(\xi)$ satisfies the Hypergeometric equation. Two linear independent solutions can be chosen as \cite{abra:00}:
 $f(\xi)=  F\left( \alpha, \beta, \gamma ; \xi \right)$ and 
  $f(\xi)= \xi^{ - 2 \rho} F\left( \alpha-\gamma + 1, \beta-\gamma +1 , 2- \gamma ; \xi \right)$;
  where   $ F\left( \alpha, \beta, \gamma ; \xi \right)$ is the Hypergeometric function. However,  the  second solution has not  the correct asymptotic behavior and has to be discarded.  The corresponding  upper component is obtained from the first equation in  (\ref{eqaco2}). The complete spinor solution is then  given as
  
    \begin{align}  \label{sol1}
\psi  =  C \, e^{ik_y y} \xi^{\rho} \, \left( 1 - \xi  \right)^{\sigma}  \fues{\begin{array}{c}
 \frac{il_B}{ \Delta d}  \, \left[    G(\xi) \,   F\left( \alpha, \beta, \gamma ; \xi \right)   
 + \xi (1 - \xi) \frac{\alpha \beta}{\gamma}  F\left( \alpha+1, \beta+1 , \gamma+1 ; \xi \right)    \right]  \\
 \\
   F\left( \alpha, \beta, \gamma ; \xi \right)  \end{array}} \, , 
\end{align}
where $  G(\xi) \, =  \,   \left[ (\rho -   2     d^2 / l^2_B )(1- \xi) -  \xi (\sigma -   2     d^2/l^2_B ) -k_yd \right] $,  
$C$ is the normalization constant and

\be \label{coef}
    \alpha = \rho + \sigma-  4 \, \left(\frac{d}{l_B}\right)^2 \, , \hskip1.5cm   \beta =  \rho + \sigma  + 4 \, \left(\frac{d}{l_B}\right)^2 + 1
      \, ,   \hskip1.5cm   \gamma  =  2 \,  \rho + 1  \, .
 \ee

\section{Bound sates}

Our aim is now to discuss the bound states spectrum. First we notice that the
 the electron-hole symmetry is preserved by the inhomogeneous magnetic field.
This follow from the fact that the  Hamiltonian in (\ref{hamilt}) 
anticommutes   with the $\sigma_3$ matrix: $\{H, \sigma_3\} = 0$.   Then, if $\psi $ is an eigenvector of $H$ with eigenvalues  $E$,    $\sigma_3 \psi $ is also an eigenvector with eigenvalue $-E$. 

\begin{figure}
\includegraphics[width=16 cm]{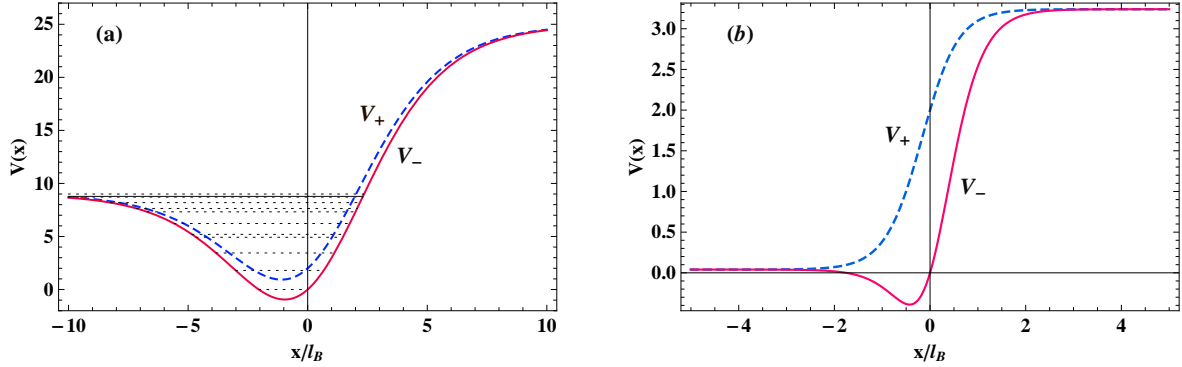}
\caption{(color online)  Plots of the effective potentials $V_{-}$ (solid lines) and  $V_{+}$ (dashed lines)  versus $x/ l_B$.  (a) The parameters  $ k_y l_B=  1 $, $d/l_B$ = 2 satisfy the condition in (\ref{cond1}). The  $\Delta^2(k_y)$  eigenvalues are represented by the  dotted lines.  (b)   $ k_y l_B=  1 $, $d/l_B = 0.4$, notice that in this case the  condition in (\ref{cond1}) is not satisfied, and therefore bounds states are not supported. }
\label{figureV(x)}
\end{figure}

The fact that the Hamiltonians   $H_+$ and $H_-$ share  the same eigenvalues, implies that  the existence of bound states   requires    that both   $V_{+} $ and  $V_{-} $    have   the form of  potential  wells. This   condition 
is obtained if the minima   of  $V_{\pm} $ are attained for a finite value of $x$, it is given by  
  \be \label{cond1}
 l_B^2 \, \vert  k_y \vert  < 2 \, d   \, .
  \ee
   As seen  in Fig.(\ref{figureV(x)}a),  when the previous condition holds   both  $V_{+} $ and  $V_{-} $ have the form of asymmetric potential  wells  around the guiding center  position 
  
 \be \label{xcl}
  x_c \, =  \,   2 d  \,  arctanh \left(  \frac{k_y }{ 2 d} \right)   \, .
\ee 
Bounds states are found if  the wells are sufficiently deep.   In the  limit in which the  transverse momentum vanishes ($k_y=0$),   the effective potentials reduce to  symmetric  wells centered at  the maximum of $B(x)$.  For values of  $k_y$ in which the condition (\ref{cond1}) is not obeyed at least one  of  $V_{\pm} $ take the form of a potential  step Fig.(\ref{figureV(x)}b), and bounds states are not supported.

The solution in Eq. (\ref{sol1}) leads to a divergence  in $\xi \to 1 (x \to - \infty)$  except for $\alpha $ or $\beta$ being a negative integer. Letting $\alpha = - n$, and utilizing Eq. (\ref{coef}) we obtain the  energy spectrum, that can be conveniently  written as 

\be \label{disprel}
  E_{n, k_y} = \pm \sqrt{2 \hbar e v_F^2 B_0} \sqrt{\left(  n - \frac{(n l_B)^2}{8d^2}\right) \left( 1 - \left(  \frac{l_B^2 k_y/2 d }{1 - n (l_B/2d)^2}\right)^2  \right)  }  \, .
\ee
The index $n$ take the values  $ n=0,1,2,3.....n_{max}$. Both  the allowed  values of   $ n $ and and $k_y$ are restricted in order to satisfy the square integrability condition, explicitly they yield

\be \label{condlig}
 n_{max}  \le  4  \, \left(\frac{d}{l_B}\right)^2   \,, \hskip1.5cm   k^d_{y ,\, max}  \le   \,  \frac{\left[4 d^2   - n l_B^2  \right]^2}{8 d^3 \, l_B^2}
  \, . 
\ee
 The first condition determines the  highest bound state supported for a given width of the barrier and 
 eliminates the possible singularity in (\ref{disprel}). Whereas, the second condition determines the allowed values of   $k_y$ for a given  $n$  and is related to   the fact that the electron   group velocity  is limited   by the free  velocity  in graphene, $i.e.$ $\vert \partial E_{n,k_y} / \partial p_y \vert  \le  v_F$. 
 The two conditions guarantee  that $ E_{n, k_y}  \, < \,  ( \hbar v_F /l_B ) \, min \, \{ V_\pm (\pm \infty)  \}$, so   the electron does not escape towards $\pm \infty$; equivalently the coefficients in (\ref{coefas}) that determine the wave function  asymptotic behavior satisfy $\rho \, , \, \sigma > 0$. 
 
  The wave function for the zero energy level takes a simple form that can be obtained by solving  the first equation in  (\ref{eqaco1}) when $\Delta=0$, it  reads

 \begin{align}  \label{sol2}
\psi  =  C \, e^{ik_y y}  e^{-k_y x}  \left[ Sech\left(  \frac{x}{2d} \right) \right]^{4 (d/l_B)^2} \,  \fues{\begin{array}{c}
0  \\
 \\
  1  \end{array}} \, . 
\end{align}
 Whereas,  for other values of $n$ the Hypergeometric functions become   Jacobi polynomials with energy dependent indices, 
the wave function is given by 

 \begin{align}    \label{sol3}
 \psi  =  C e^{ik_y y} (1+z)^{\rho} \, \left( 1 - z  \right)^{\sigma}  \,    \fues{\begin{array}{c}
 \frac{il_B}{ 2\Delta d}  \,   \left[M(z) \,   
   P_n^{(\gamma -1,-n+\beta -\gamma )}(z)    
 -  (1 - z^2) \,  \frac{\alpha \beta}{2n}  \,   P_n^{(\gamma -1,1-n+\beta -\gamma )} (z)   \right]   \\
 \\
    P_n^{(\gamma -1,-n+\beta -\gamma )}(z)   \end{array}} \, ,
\end{align}
where $z=2\xi -1 =  tanh(x/2)$ and $M(z) \, =(\rho -   2     d^2/l^2_B )(1- z) -  (\sigma -   2     d^2/l^2_B )(1+z) - 2 k_yd $.

\begin{figure}
\includegraphics[width=10cm]{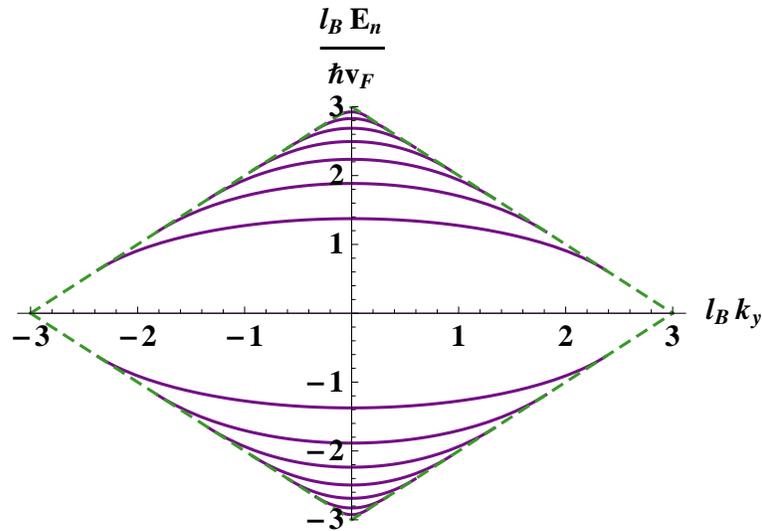}
\caption{(color online)    Bound states energy spectrum $E_n(k_y)$ as a function of $k_y$ for a magnetic barrier with $d = 1.5 \, l_B$.  According to   (\ref{condlig})  
$n_{max}=9 $. For every $n-$level  the allowed values of $k_y$  are delimited by the free-electron spectrum  $E=\pm \hbar v_F k_y$ (dashed lines).   }
\label{especk}
\end{figure}

The dispersion relation in (\ref{disprel})  shows  that the inhomogeneity of $B$ lifts the degeneracy for  every quantum level $n$
and gives rise to a $k_y$-dependent dispersion relation, which leads to a drift velocity along the $y$ axis. 
 This is valid, except for the $n=0$ level that has zero energy, independent on the magnetic field for all  values of $k_y$.
 The energy spectrum
for $d =  1.5  l_B$  is displayed  in Fig.(\ref{especk}) as a function of $k_y$. For the selected values we have   $n_{max}=  9  $, and each level $n$ results in a band as the transverse momentum sweeps  from $k_y=0$  to the maximum value  $k_{y,max}^d$ in (\ref{condlig}).

\begin{figure}
\includegraphics[width=10cm]{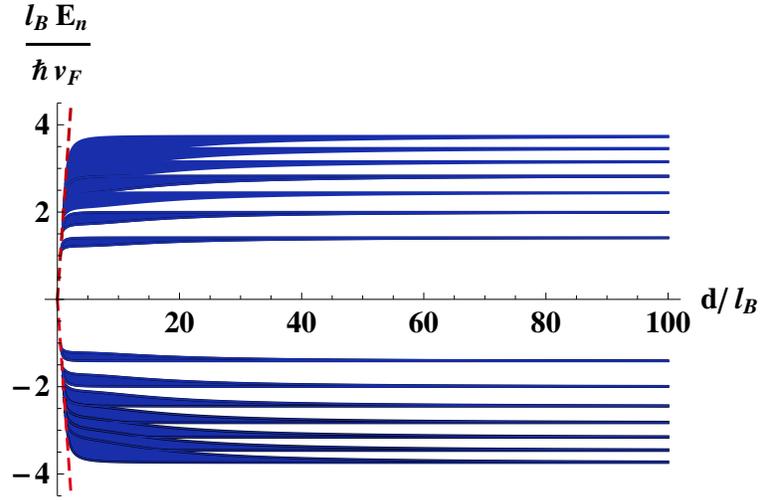}
\caption{(color online)    Energy spectrum  as a  function of $d$ for a magnetic barrier when the system is confined by a square box of area $L  \times L$ with $L = 50 l_B$. The dark zones are the allowed energy values.  
The dashed line  $E l_B /\hbar v_F = 2 d /l_B$ delimitates the  minimum  barrier width that supports bound states. 
For every level  $n$ the transverse momentum  varies between $k_y=0$
and   $k_y= K_{y, \, max}$.   }
\label{especd}
\end{figure}

The previous results apply when the width of the  barrier is small in comparison with the  system dimensions.  We  now analyze the behavior of the spectrum  as we change from a narrow to a broad barrier, considering that the system is confined within a square box of length 
$L \times L$. 
For a narrow barrier the values of $k_y$ are limited by the second equation in (\ref{condlig}). Instead, when the size of the barrier is comparable to $L$, the number of allowed states is limited by those  that can be accommodated in the square box. Assuming  periodic conditions   for the wave function in (\ref{sol3})   along  the 
$y-$direction, yield $ k_y = 2 \pi \, j / L $ with $j$ an integer. But  according to Eq. (\ref{xcl})  $k_y$ also determines the center position  $x_c$  of the electron, hence  $x_c < L/2 $ and   the number of quantum states is given  by $N= j_{max} = ( 2 d L /\pi l_B^2)   \tanh\left( L/4 d \right)$, whereas the momentum  limit imposed  by the size of the system reads 
    $k^L_{y, \,  max} = 2(  d/l_B^2) \tanh\left( L/4 d \right) $.  It is interesting  to notice,  that similarly to the homogeneous case,  the degeneracy can be written as
 $N  = \Phi / \Phi_0$, where the magnetic flux  produced by  the field   in  (\ref{magB}) through the sample   is 
   $ \Phi =  4 B_0   d L     \tanh\left( L/4 d \right)$ and  $ \Phi_0 = h/e$ is the elementary fluxon.  In order to track the degeneracy evolution as the barrier is modified from  narrow to broad as compared to $L$, we  define
   
   \be \label{cutk}
  \frac{1}{K_{y, \, max} } =  \frac{1}{k^L_{y, \,  max} } +  \frac{1}{ k^d_{y, \, max }}    \,  
\ee
  as a  cut for the transverse momentum. Notice that  $K_{y, \, max}$  interpolates between    $k^d_{y \, max} $   valid for a narrow barrier,  and $k^L_{y, \,  max}$ 
   valid when $ d > L$. 
 Fig.(\ref{especd})  shows the resulting energy spectrum, the dark zones are the allowed energy values. For every level  $n$ the transverse momentum  varies between $k_y=0$
and   $k_y= K_{y, \, max}$.  The restriction on the level index $n$ given by the first equation in (\ref{condlig}),  translates into 
the following equation for the separatrix $E = 2 \hbar v_F d /l^2_B$ (dashed line),    energies  to the left 
of this line  are not allowed.  For small value of $d/l_B$  a few  bands  and gaps can be identified for the first values of $n$, as the energy is increased we observe a  continuous energy region up to the maximum allowed value
$n_{max}$.  As the size of the barrier increases  a larger  number  
of  energy gaps appear.   When the size of the magnetic barrier becomes comparable to $L$ the width of the  bands decrease. Finally  in the 
 homogeneous $B$   limit, $d \gg  L$, the energy eigenvalues reduce to the relativistic massless Landau levels  observed in graphene: $   E_{n,k_y} = \pm \sqrt{2 \hbar e v_F^2 B_0 \, n}$,  independently of $k_y$, whereas the degeneracy reduces to the well known  result $N = \Phi / \Phi_0$, with  $ \Phi = B_0 L^2$.

\section{Scattering}

We now consider the scattering regime. A  plane wave incident from  $ x \to - \infty \, , (\xi \to 1)$ propagates at an angle $\phi$ with respect to the $x$ axis. Taking into account the  gauge selection  in (\ref{magA}) the incoming   momenta are  parametrized  as

\be \label{defmom}
l_B \,  k_x =  \Delta   \cos \phi \, , \hskip1.5cm  l_B \, k_y =  \Delta  \sin  \phi  + \frac{2 d}{l_B}  \,  .
\ee
 The transmitted  wave has a longitudinal momentum 
  $k_x^\prime  = \frac{\Delta}{l_B} \cos \phi^\prime $
 where $\phi^\prime$  is the  refracted angle. The conservation of $k_y$ gives the relation between the  
 incident and refracted angle as follows   

 \be \label{relang}
 \sin  \phi^\prime    =  \sin  \phi  + \frac{4 \,   d }{ l_B \, \Delta} \, , 
 \ee
 whereas energy conservation allow us to relate the transmitted  and incident longitudinal momenta as
 
\be \label{relmom}
 l_B \, k_x^\prime  = \sqrt{\left( l_B \, k_x \right) ^2 - \  \left( \frac{4 d}{l_B}  \right)^2 - 8 k_y d   } \, . 
 \ee 
 Eq. (\ref{relang})  implies that for a critical  angle $\phi_c = \arcsin \left(1 - 4 d/l_B \Delta \right) $ no transmission  is possible.  Furthermore,  when the following condition applies
 
 \be \label{condlim}
  \frac{E l_B}{\hbar v_F}   \, \le \,2\,  \frac{d}{l_B}  \, , 
 \ee
the transmission vanishes regardless of the incident angle $\phi$.   The   condition in  (\ref{condlim}) 
  establish  that states with an average cyclotron radius  (in the barrier region) 
smaller that $2 d$ will bend by the magnetic field and are completely  reflected, a similar condition was obtained in the case of a square-well magnetic barrier
\cite{mart:066802}.
 Comparing  the  equations (\ref{defmom})  with (\ref{coefas}), we observe that  the 
 longitudinal momenta $k_x$ and $k_x^\prime $  are related to the asymptotic coefficients $\rho$ and 
 $\sigma$   as follows: $\sigma = - i  k_x \, d $ and 
 $\rho =  d \,  \sqrt{ -k_x^{\prime2}   }$.  It is verified that if condition  (\ref{condlim})  holds $\rho$ is real, instead when (\ref{condlim})  is not valid we replace  $\rho =  - i  k_x^\prime \, d$.
    Utilizing the properties of the Hypergeometric functions it is  verified  that  in the limit $x \to \infty \, (\xi \to 0)$  the 
wave function in (\ref{sol1}) yields the correct asymptotic expression 

 \begin{align}  \label{asymp1}
\psi  \sim    \, e^{i(k_x^\prime x + k_y y)}  \,   \fues{\begin{array}{c}
s \, e^{-i \phi^\prime} \\
 \\
  1  \end{array}} \, , 
\end{align}
where $ s = sgn \, E$. 
 The asymptotic value of the  wave function for  $ x \to - \infty \,  (\xi \to 1)$, is obtained using the linear transformation  formulas\cite{abra:00}  that relate 
 $F(\alpha,\beta,\gamma, \xi)$ with Hypergeometric functions evaluated at $1-\xi$, 
 to obtain 
 
\begin{align}  \label{asymp2}
\psi  \sim    \, e^{i(k_x  x + k_y y)}  \, 
\frac{\Gamma \left(\gamma \right) \Gamma\left(\gamma - \alpha - \beta \right) }{\Gamma\left(\gamma -\alpha \right) \Gamma\left(\gamma  - \beta \right) }
\,   \fues{\begin{array}{c}
s \, e^{i \phi  } \\
 \\
  1  \end{array}}    
  \, \, + \,\, 
   \, e^{i(- k_x  x+ k_y y)}  \, 
   \frac{\Gamma \left(\gamma \right) \Gamma\left( \alpha +  \beta - \gamma  \right) }{\Gamma\left( \alpha \right) \Gamma\left(  \beta \right) }
     \fues{\begin{array}{c}
- s \, e^{-i \phi  } \\
 \\
  1  \end{array}} \, .
\end{align}
  
  From this equation the reflection coefficient $R$  is obtained as 
  
    \be \label{coefR}
  R=   \Bigg\vert  \frac{\Gamma \left(\rho - i k_x d + (2d/l_B)^2+ 1  \right) \Gamma\left(\rho - i k_x d -  (2d/l_B)^2  \right) }{\Gamma\left( \rho +  i k_x d + (2d/l_B)^2+ 1 \right) \Gamma\left( \rho  + i k_x d -  (2d/l_B)^2  \right) }   \Bigg\vert^2 \, .
  \ee 
Under  condition (\ref{condlim})  $\rho $ is real, thus   $R=1$  and,  as expected, the transmission coefficient vanishes. Instead, if  the condition (\ref{condlim})  is not obeyed, $\rho $  is substituted 
by   $\rho = - i d k_x^\prime $, and the  transmission probability  $T = 1 - R$  can be written in a simple closed form as

 \be \label{coefT}
 T \, = \,   \frac{\sinh \left[2\pi l_B k_x  \right] \, \sinh \left[2\pi l_B k_x^\prime \right] }{ \sin^2 \left[4\pi \left( \frac{d}{l_B}\right)^2 \right] \, +\,  \sinh^2 \left[\pi l_B ( k_x + k_x^\prime)  \right]} \, . 
 \ee

\begin{figure}
\includegraphics[width=5cm]{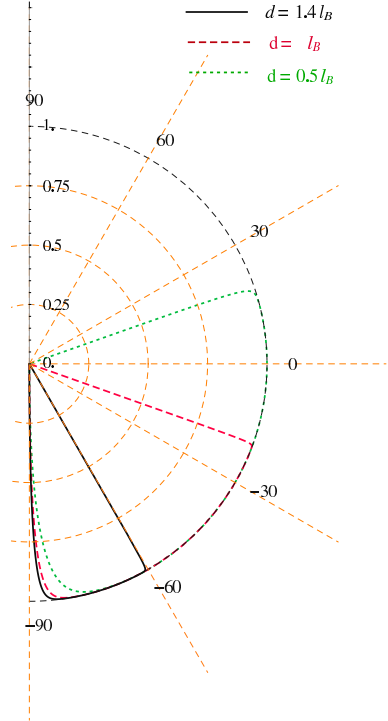}
\caption{(color online)   Angular dependence of the transmission coefficient $T$ through a barrier  with energy 
$\Delta = E l_B/\hbar v_F = 3  $ and  different  values for $d$:
$d=1.4 l_B$ (continuos line), $d=  l_B$ (dashed line), and $d=0.5 l_B  $ (dotted line).
   }
\label{PolaPlot1}
\end{figure}

\begin{figure}
\includegraphics[width=12cm]{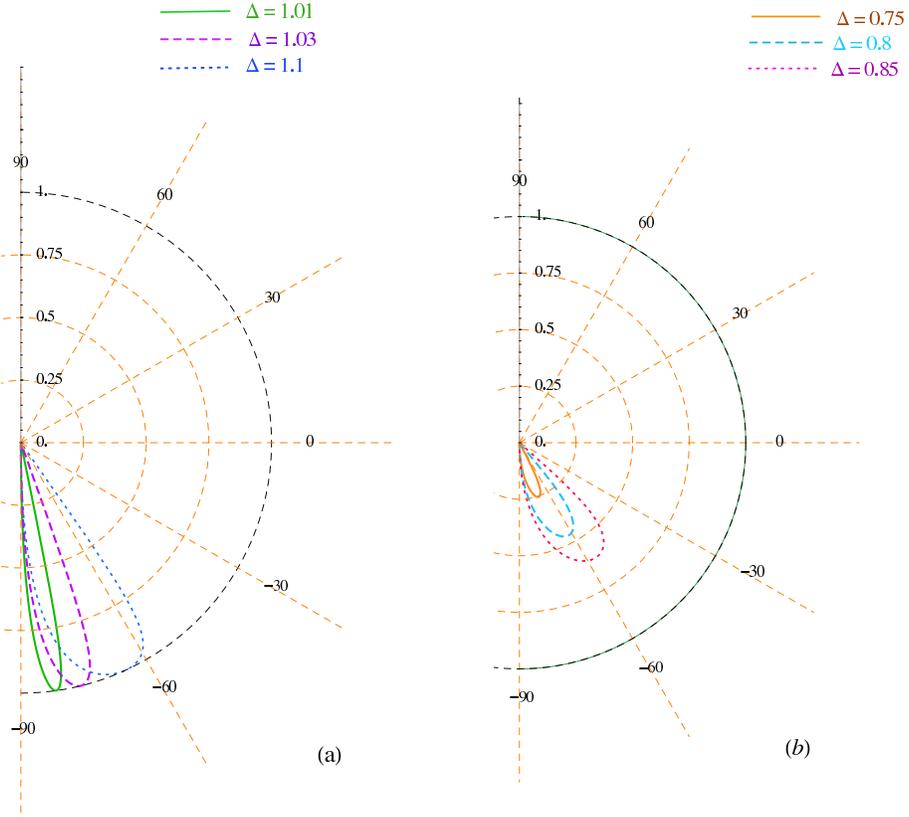}
\caption{(color online)  (a)   Angular dependence of the transmission coefficient $T$  through a barrier  with
 $d = l_B/2$ (resonant condition in Eq. (\ref{resonTmax})  with $j=1$), 
 and the following  values of the energy:
$\Delta= 1.01$ continuos line, $\Delta=1.03 $ dashed line, and $\Delta=1.1 $ dotted line.
(b) As in (a) for $d = l_B/\sqrt{8}$ (Eq. (\ref{resonTmin})  with $j=0$)
 and the following  values of the energy:
$\Delta= 0.75$ (continuos line), $\Delta= 0.8$ (dashed line), and $\Delta= 0.85$ (dotted line).
 }
\label{PolaPlot2}
\end{figure}

  A  plot for the transmission coefficient as a function of the incidence angle  is shown in 
  Fig.(\ref{PolaPlot1})    for a fixed energy  and  several values of  $d$.  The qualitative behavior in this plot 
  is similar to  those obtained for the case of a square well magnetic barrier \cite{mart:066802}.
  However an advantage of the 
 the result  in (\ref{coefT})  is that it allow us to identify the resonant conditions at which  the barrier becomes transparent  ($T \sim 1$), it is given by 
  
   \be \label{resonTmax}
d = \frac{\sqrt{j}}{2} \, l_B \, \hskip1.5cm j= 1,  2 ,3,.. ... 
 \ee
   In this case the barrier acts as an asymmetric  filter, it behaves as  perfectly transparent  for angles
   in the region $ -\pi/2 < \phi <   \phi_c $  Fig.(\ref{PolaPlot2}a).  In particular for  energy values 
   slightly above the threshold condition in  (\ref{condlim}),  $ \Delta = d/l_B + \epsilon $, with  
    $ \epsilon \ll 1 $,  the  width of the transparency region 
   can be very narrow. 
   
\begin{figure}
\includegraphics[width=18cm]{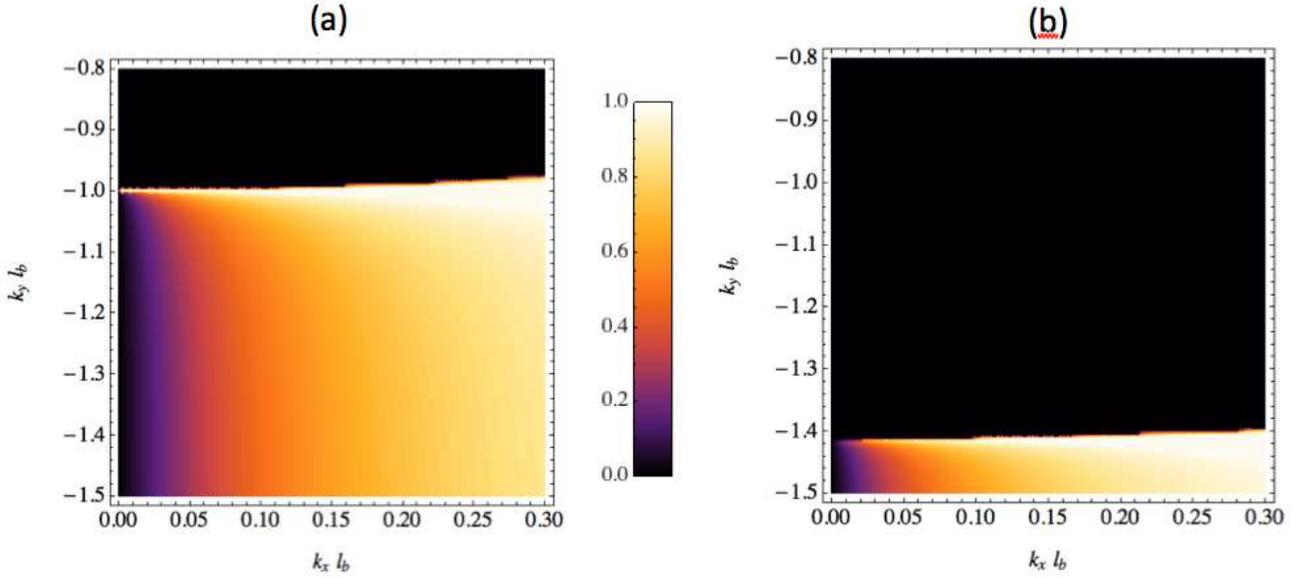}
\caption{(color online)  a)   Contour plot of  the transmission coefficient $T$  through a barrier  with
 $d = l_B/2$ (resonant condition in Eq. (\ref{resonTmax})  with $j=1$), 
 (b) As in (a) for $d = l_B/\sqrt{8}$ (Eq. (\ref{resonTmin})  with $j=0$).
 }
\label{ContourPlot}
\end{figure}

   On the other hand, under the condition 
     \be \label{resonTmin}
d = \frac{\sqrt{j +\frac{1}{2}}}{2} \, l_B \, \hskip1.5cm j=0,1,  2 ,3,.. ... 
 \ee
 the value of $T$ is reduced, in particular for values of the energy 
  slightly    above the threshold condition in  (\ref{condlim}), $ \Delta = d/l_B + \epsilon $, with  
    $ \epsilon \ll 1 $, the transmission coefficient is  strongly  reduced, Fig.(\ref{PolaPlot2}b).
   Another way to contrast these results  is shown in   Fig.(\ref{ContourPlot}) where a contour plot for the transmission coefficient $T$ is shown. A can  be seen 
    in Fig.(\ref{ContourPlot}(a)  corresponding to  the  resonant condition in Eq. (\ref{resonTmax})   there is a wider region  for the transmission that is not present in Fig.(\ref{ContourPlot}(b)  corresponding to  the  minima  condition in Eq. (\ref{resonTmin}).

  \section{Conclusions}
In conclusion, we consider  the dynamics  of carriers in  graphene  subjected to an inhomogeneous hyperbolic  magnetic field.  The corresponding Dirac equations was analyzed 
 within the formalism of supersymmetric quantum mechanics. We found compact analytical solutions for the energy eigenvalues and eigenfunctions for electrons and holes. 
 The dispersion relation in (\ref{disprel})  shows  that the inhomogeneity of $B$ lifts the degeneracy for  every quantum level $n$
and gives rise to a $k_y$-dependent dispersion relation, which leads to a drift velocity along the $y$ axis. 
 This is valid, except for the $n=0$ level that has zero energy, independent on the magnetic field for all  values of $k_y$.
  For a narrow barrier the spectra displays a series of bands separated by gaps, as the  width of the barrier increases we can track  the levels evolution into the degenerated Landau levels. In the scattering  regime  a simple analytical formula is obtained for the 
transmission coefficient,  this result  allow us to identify the resonant conditions at which  the barrier becomes transparent.  
 We expect that the  results obtained in this work will  be useful in order to analyze the confinement by  magnetic barriers in graphene samples. In a future work we plan to address the problem of calculating the longitudinal conductivity as well as the Hall conductivity for the case of graphene under 
 inhomogeneous magnetic fields.

\acknowledgments
 We acknowledge support 
from UNAM project PAPIIT IN118610.

 

\newpage

 \end{document}